\documentclass[12pt,aps,onecolumn,nobibnotes]{revtex4}
\usepackage{graphicx}%
\usepackage{dcolumn}
\usepackage{amsmath}
\usepackage{amssymb}

\def\be{\begin{equation}}
\def\ee{\end{equation}}
\def\bea{\begin{eqnarray}}
\def\eea{\end{eqnarray}}
\def\nn{\nonumber \\}
\def\openone{\mbox{1\kern -0.25em I}}

\def\openZ{\mathbb{Z}} 
\def\openN{\mathbb{N}}

\def\Cl{C\kern -0.2em\ell}

\def\ed{\end{document}}
\def\Ext{\text{Ext}}

\def\d{\text{d}}
\def\dwedge{\,\dot\wedge\,}
\def\JJ{\,\rule{5pt}{1pt}\rule{1pt}{6pt}\,}
\begin{document}

\title[Quantum Clifford Hopf gebra for QFT]{%
QUANTUM CLIFFORD HOPF GEBRA\\
FOR QUANTUM FIELD THEORY}
% Force line breaks with \\
\author{Bertfried Fauser}
\email{Bertfried.Fauser@uni-konstanz.de}
\affiliation{% 
Fachbereich Physik\\
Universit\"at Konstanz, Fach M 678\\ 
78457 Konstanz, Germany\\ 
}
\begin{abstract} 
We give arguments for the necessity to employ quantum Clifford Hopf gebras 
in quantum field theory. The role of the antipode is examined, Feynman
diagrams are re-interpreted as tangles of graphical calculus. Regularization
due to the design of convolution Hopf gebras is given as a program for future 
research.
%17B37 Quantum groups
%15A66 Clifford algebras, spinors
%11E39 Bilinear and Hermitian forms
algebra, Quantum field theory, renormalization, geometric algebra
\end{abstract} 
\maketitle 

\vspace{-1.0truecm}
\section{Introduction}
Clifford algebras have been used to code QFT for several times, e.g.
\cite{Caianiello}. Our starting point is `functional QFT' developed
by Stumpf and coworkers \cite{StuBor}. In this approach quantum field 
theories are described by algebraic sources (Schwinger sources) spanning
a functional (Fock) space. The coefficients of such sources are correlation
functions w.r.t. the exact physical vacuum, as e.g. time-ordered 
$\tau$--functions or normal-ordered $\phi$--functions. The dynamics is 
given via the Schwinger--Dyson hierarchy which translates into a functional 
equation in the algebraic picture, see (\ref{eq:5}). Using methods closely 
related to $C^*$-algebras like GNS-states etc. the hierarchy
translates into a functional Schr\"odinger equation which contains the
whole information of the quantum dynamics. Unfortunately calculations
in this formalism become cumbersome, if compared with e.g. path integral 
methods, which prevented their common usage, in spite of successful
applications of the method in composite particle theory and the 
successful derivation of the $SU(1)\times SU(2)$ electroweak theory 
from sub-fermion models which are beyond the abilities of path integral 
methods \cite{StuBor}.

This computational success was the motivation to search for a compact 
formalism behind the method. In several works 
\cite{F-vertex,F-positronium,F-thesis,F-transition,F-wick}
a concise and powerful description of functional quantum field theory 
was developed employing Clifford algebras. However, it turned out that
one had to use `Quantum Clifford Algebras' (QCA) \cite{FauAbl}, i.e. Clifford
algebras of an arbitrary bilinear form, which cover deformed structures
as Hecke algebras, the Manin quantum plane, $q$-spinors etc. The first success 
of this treatment was the disappearance of certain singularities in 
vertex normal-ordering \cite{F-vertex} just by a correct algebraic
reformulation. A further point concerns interacting QFT which cannot be 
treated in Fock space, i.e. with path integral methods. Therefor such QFT 
has to {\it select} a unique vacuum for every dynamics \cite{F-vacua}. 
It was shown, that this is the case if the antisymmetric part of the 
bilinear form of the QCA is identified with the (exact) propagator of the 
theory. Perturbation theory uses the free propagator and thus falsely
Fock space.

In this paper, we give arguments, that a `Quantum Clifford Hopf Gebra`
(QCHG) should be used to describe QFT. The most important technique is
Rota-Stein `Cliffordization' \cite{RotSte}, which allows to introduce
Clifford products by product `mutation' of the Gra{\ss}mann product
and yields {\it closed} and {\it grade independent} formulas. This process
needs intrinsically the co-gebra structure.

The convolution product given canonically by a pair of product and 
co-product and the antipode constitute a Hopf gebra 
\cite{Ozi-CJP,Cruz}. 
Product, co-product and Convolution will be interpreted by quantum 
processes like absorption, emission and interaction \cite{Ozi-ixtapa}. 
The meaning of the antipode can be anticipated from \cite{F-wick}, 
where it was identified with grade-involution in the Gra{\ss}mann case.
The antipode plays also an important role in renormalization theory
of perturbative QFT \cite{Kreimer}. However, as discussed above, we
are searching for non-perturbative regular QFTs. In this spirit, we
state a program how a Hopf-gebra could be designed for such a purpose. 
\section{Quantum Clifford Algebras for QFT}
The definition of QCAs can be found in \cite{Ozi-FGTC,FauAbl}. We give
the basic definitions to show how this algebras fit for QFT. Using 
Chevalley deformation (see the criticism below) we define 
($x,y \in V$, $u,v,w \in V^\wedge \equiv \Ext V$)
\begin{eqnarray}
\begin{array}{crcl}
i)  & x \JJ_B y & = & B(x,y) \\
ii) & x \JJ_B (u \wedge v) & = & (x \JJ_B u) \wedge v
                              - S(u) \wedge (x \JJ_B y) \\
iii)& u \JJ_B (v \JJ_B w) & = & (u \wedge v) \JJ_B w \\
vi) & S & : & V^\wedge \mapsto V^\wedge, 
            \quad\quad S(u) = (-1)^{\partial u}u,
\end{array}
\label{eq:1}
\end{eqnarray}
where $\partial u \in \openN_0$ is the Gra{\ss}mann grade of $u$. It is obvious
from this definitions that the bilinear form $B$ needs to have no 
symmetry, while the usual commutator based definition $\{x,y\}_+ = 2g(x,y)$
is by definition restricted to symmetric bilinear forms. We define
$g(x,y):= 1/2(B(x,y)+B(y,x))$ and $F(x,y):=1/2(B(x,y)-B(y,x))$.
There is an Clifford {\it algebra} isomorphism $\phi$, called 
Wick-isomorphism \cite{FauAbl}, which connects $\Cl(V,B)$ and $\Cl(V,g)$. 
This results in a change of the Gra{\ss}mann product and of the 
Gra{\ss}mann grade. One defines
\begin{eqnarray}
\label{eq:4}
x \dwedge y &:=& x \wedge y + F(x,y)
\end{eqnarray}
on grade one elements which is obviously not grade preserving, i.e.
$\partial \not= \dot{\partial}$, \cite{FauAbl,F-wick}. 

Applying this structure to QFT is done by the following identification.
Let $\psi_X$ be a field operator (or its dual) with all indices 
(discrete, continuous and conjugation) are put into a super-index $X$. 
The field operators are identified with the Clifford map as
\cite{F-dirac,F-transition}:
\begin{eqnarray}
\label{eq:2}
\psi^L_X := X \JJ_g + X \wedge + F(X,\ldots) \\
\label{eq:3}
\psi^R_X := X \JJ_g - X \wedge + F(X,\ldots) \, ,
\end{eqnarray}
where L,R denotes right or left action. Given a possibly non-linear
field equation $\dot\psi = V(\psi)$ we obtain the normal-ordered
generating functional $\mid N(j,a)\rangle$ in Clifford terms via 
\cite{F-transition}
\begin{eqnarray}
\dot{\psi} &=& [H(\psi),\psi]_- \nn
H[j,\d] &=& H(\psi^L)\psi - \psi H(\psi^R) \nn
\partial_t \mid N(j,a)\rangle &=& H[j,\d] \mid N(j,a)\rangle \, ,
\label{eq:5}
\end{eqnarray}
where $j$ and $\d$ algebraically span $V^\wedge$ and its dual 
$V^{*\wedge}$. One notes: Quantization is given by the symmetric part 
of the bilinear form, while the propagator is given by the antisymmetric 
part which fixes all freedom.
\section{Bigebra Structure}
Every space $V^\wedge$, $\dim V=n$ underlying a Gra{\ss}mann algebra 
has a $\openZ_n$-grading, and there is a natural duality between 
grade $r$ and $n$-$r$ subspaces, e.g. $V$ and $\wedge^{n-1}V$. Beside the 
deformation into a Clifford algebra as in (\ref{eq:1}), one can think 
about introducing a Clifford product on $\wedge^{n-1}V$ \cite{Con}. 
Using product co-product duality (see Fig. \ref{prod-coprod}), it is well 
know, that any product on the dual space induces a co-product on the 
former. A Gra{\ss}mann algebra is in this way a {\it convolution algebra} 
obeying algebra and co-gebra structure. However, it can be proved that it 
is a bigebra possessing an antipode S (given in (\ref{eq:1}-vi) !), 
rendering it a Hopf gebra. If an unital convolution algebra possess 
an antipode it can be proved to be a Hopf gebra \cite{Ozi-ixtapa,Cruz}. 
We re-interpret the product as an {\it absorption} (of a tensor factor) 
and co-product as an {\it emission} (of a tensor factor).
\section{Clifford Hopf gebra}
Since a Gra{\ss}mann Hopf gebra is the classical case, one is
interested to find a Clifford Hopf gebra as its quantization. 
Unfortunately, there are theorems \cite{RotSte,Ozi-CJP} that 
Clifford convolution algebras possess {\it no} antipode if product 
co-product duality, see Fig. \ref{prod-coprod}, is employed. 
Moreover in \cite{RotSte} only the product was deformed, which is 
artificial.
\subsection{Convolution Bigebra}
A program was started \cite{Ozi-CJP} which breaks the product co-product
duality and investigates the general case of a Clifford convolution 
algebra. However, since the antipode seems to play an important role in 
renormalization theory \cite{Kreimer} and might be needed to model
some notions of physics like expectation values, and the `split' of
a tangle, see below, we want to propose a different approach. 
\subsection{Antipode}
\begin{figure}[t]
\parbox[t]{0.50\textwidth}{\hspace{-1.75truecm}
\includegraphics[height=1.5truecm]{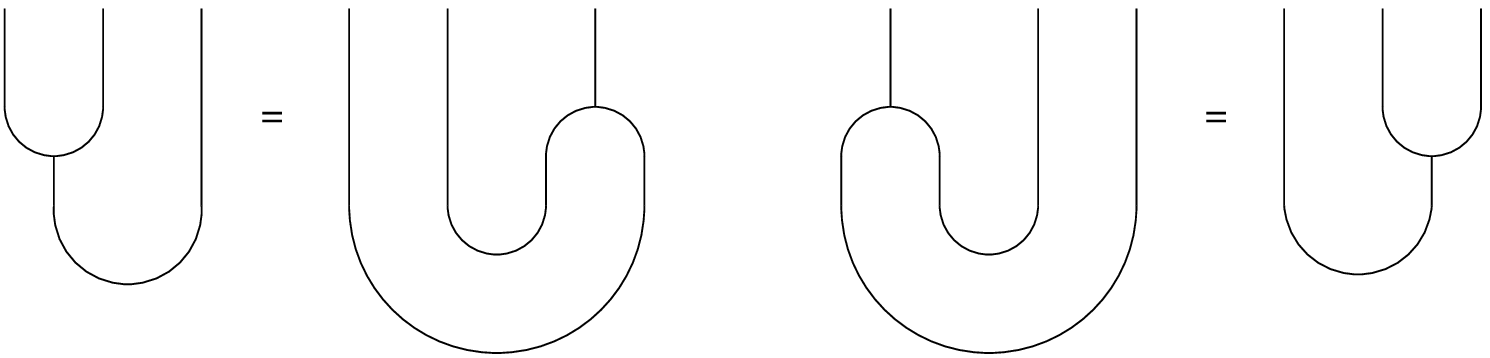}
\caption{Product Co-Product Duality}
\label{prod-coprod}
}
\hfill
\parbox[t]{0.40\textwidth}{\hspace{-1.75truecm}
\includegraphics[height=1.5truecm]{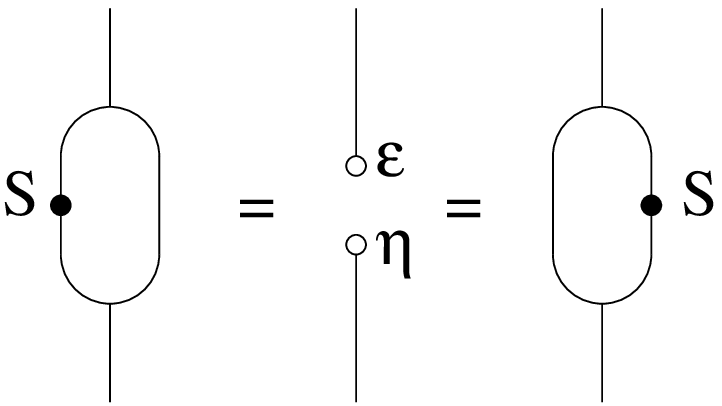}
\caption{Antipode Axioms}
\label{antipode}
}
\end{figure}

Axioms for an antipode can be given for any unital convolution algebra,
see Fig. \ref{antipode}.
If an algebra is unital and a co-gebra is co-unital, the mutually defined
convolution algebra is unital too. If an antipode can be found for a given 
pair of product and co-product, it is unique. A quadruple $H=(V^\wedge,
m,\Delta,S)$ of a space $V^\wedge$, a product $m$, a coproduct $\Delta$
and an antipode $S$ is a Hopf gebra \cite{Ozi-ixtapa,Cruz}.
\subsection{Quantum Clifford Hopf gebras}
An idea developed in \cite{FauOzi} to circumvent the non-existence theorem
for the antipode is to employ QCA. Having an antisymmetric part in the
bilinear form it is possible to arrange that det$(g)\not=0$ but 
det$(B)=0$. In this case, no product co-product duality is possible and 
we can freely adjust product and co-product. Indeed antipodes for $1$+$1$
space-time have been found in such a setting. 
\section{QFT by Clifford Hopf Gebras}
Quantization in the sense of a transition from Gra{\ss}mann to Clifford
algebras has a major drawback if introduced by Chevalley deformation
as done in (\ref{eq:1}). The first two formulas (\ref{eq:1}-i) and 
(\ref{eq:1}-ii) are applicable {\it only} for grade one elements. This
means, that we have to fix a unique grade one space, which breaks some 
symmetries. Moreover, computations with higher grade elements have to be
reduced recursively to this cases. {\it But} -- this is also exactly
the case in QFT if described in the `second quantization' picture, or even
in quantum mechanics if the Dirac creator, annihilator picture is used. 

Creation is (wedge) concatenation by a grade one 
element, while annihilation is the anti-derivation rule given in 
(\ref{eq:1}-ii). This is reflected by formulas (\ref{eq:2}) and
(\ref{eq:3}) where we identified the field operators. {\it But} --
as we have seen in (\ref{eq:4}), adding an antisymmetric part does
{\it not} preserve the grading. This is the (Hopf algebraic) origin
of the need of a Wick normal-ordering which literally does transform
to that particular grading which is {\it selected} by the propagator 
(antisymmetric part) \cite{F-wick}.
\begin{figure}[t]
\parbox[t]{0.58\textwidth}{\hspace{-2.25truecm}
\includegraphics[height=1.5truecm]{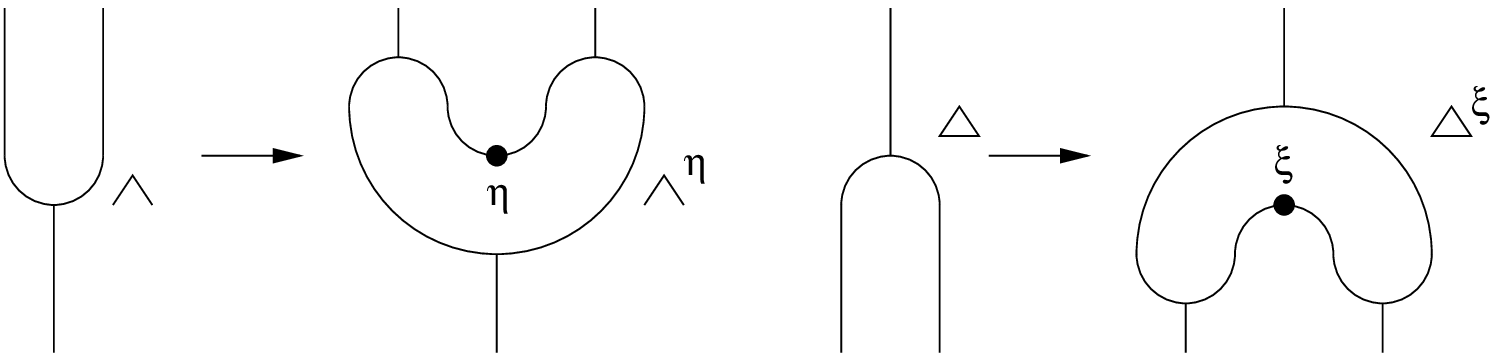}
\caption{Bicliffordization: \newline The Rota--Stein `sausage' tangle.}
\label{cliffordization}
}\hfill
\parbox[t]{0.40\textwidth}{\hspace{-1.75truecm}
\includegraphics[height=1.5truecm]{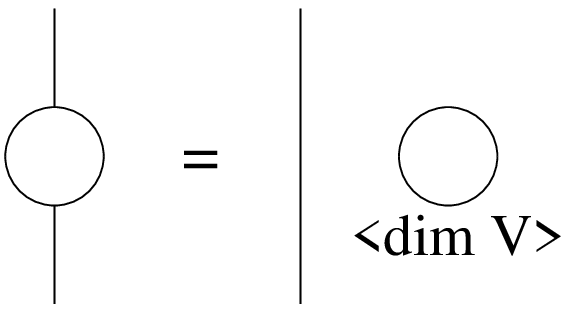}
\caption{Loop in Gra{\ss}mann or Clifford convolution with $B^*=B^{-1}$}
\label{loop}
}
\end{figure}
Using Rota--Stein's `Cliffordization' \cite{RotSte}, one can introduce a
Clifford product as a deformation of the Gra{\ss}mann product for any
bigebra element. This deformation, shown in Fig. \ref{cliffordization},
replaces the product tangle by a more complicated tangle, possessing an 
internal  loop. Since it has two inputs and one output, it is still a product
or an absorption. The same should be done with the co-product, by duality 
as shown in Fig. \ref{prod-coprod}, or starting from a Clifford convolution 
algebra with two independent (might be singular) bilinear forms. Having 
such a general tool at our disposal one can break of the strong connection 
which ties grade one elements to absorption and emission or say annihilation 
and creation.  
\subsection{Feynman Diagrams as Tangles}
We re-interpret, as also promoted by Oziewicz \cite{Ozi-ixtapa}, 
a convolution bigebra tangle as physical process.
This is quite different to the usual interpretation of such iconic 
notations. Perturbative QFT (pQFT) creates and annihilates single
bare --or naked-- particles or fields. Only after this unphysical 
processes has been defined the particles are {\it dressed} by an infinity 
set of possible internal {\it vacuum polarization} diagrams. However 
this method can be see to follow directly from the drawback of the 
definitions in Eqn. (\ref{eq:1}). Since only grade one elements can be used 
in the creation and annihilation process, any higher grade action has to be 
split into this basic building blocks. Such a behaviour can be found e.g.
in Wick normal-ordering \cite{F-vertex,F-wick}. If the base space $V$ is
of infinite dimension, such an process is an infinite recursion.
The solution may be taken from the Rota--Stein Cliffordization. Since this
formula is grade independent, due to the usage of Hopf algebraic methods,
such tangles are valid for {\it any process}, not just only creation and
annihilation by grade one elements. This opens very general new insights
into the structure of non-perturbative QFT. pQFT is moreover tied to the
intrinsic grading obtained from the artificial distinction of grade one
elements. A change to this grading e.g. by Wick normal-ordering, 
{\it has to be performed}, otherwise no numerical results can be 
obtained. The Hopf gebraic approach cannot suffer from this deficiency,
since it is intrinsically invariant w.r.t. the grading.

The Rota-Stein Cliffordization depicted as `sausage' in Fig.
\ref{cliffordization} shows a second feature which is of utmost importance 
for QFT. Adding a certain {\it internal} loop, mutates a product into 
a new one. Indeed, the tangles before and after mutation (i.e. 
quantization) work out completely different. This can be the starting 
point for an axiomatization of Hopf gebraic structures of convolution 
algebras of QFT. Of course this is a wide and open field of 
speculation, but looking carefully at the needs of QFT will surely 
generate such rules. The main problem will be to find some basic
tangles, the processes of QFT, which may possess internal loops. Such 
basic tangles have to be selfconsitent. Afterwards they can be easily 
tied together to model interactions. In the case of the Clifford product 
this is already a truth. One should however be careful to think too 
simple minded. It will be most likely the case, that not a finite set
of `basic' tangles will do the job in any case. When proceeding from 
elementary objects to bound states, one should expect a new {\it basic} 
behaviour of such objects resulting in the need of new tangles with 
internal loops. `Elementary' tangles might show an internal structure
if `magnified' like the Clifford product if seen as sausage tangle.
\subsection{Renormalization}
We discuss how such techniques are employed in the theory of renormalization. 
Connes, Kreimer and others have used the antipode of a rooted tree Hopf 
gebra \cite{GL} to generate the counterterms of renormalization 
\cite{Kreimer}. This is analogous to the Wick normal-ordering using Hopf 
gebra techniques \cite{F-wick}. However, both cases fail to hit the 
point, since we are interested in invariants. E.g. the artificial grade
one dependence forcing normal-ordering is also artificial. The same 
argument should hold for renormalization, which is also an artificial 
rewriting of diagrams into a series of diagrams which have to 
be summed up. Such diagrams are {\it not} tangles in our sense!
\subsection{Search for Finite QFT}
We end this article proposing a program to search for a structure
which may, this is our hope, lead ultimatively to finite QFT. 

Suppose you have a loop as given in the LHS of Fig. \ref{loop}. 
If such a tangle is seen as physical process it cannot be distinguished
from a single line. Otherwise one would have to have information about 
the internal structure of the tangle, but that would need a measurement 
and this would incorporate a different process as mere propagation.

If we calculate this tangle in a Clifford biconvolution bigebra 
\cite{Ozi-ixtapa,Cruz,FauOzi}, i.e. with the requirement that the 
product and coproduct is Clifford and they are related by $B = B^{-1}$, 
we get the RHS. We stay with a single line up to a scalar factor which 
equals the dimension of the space (infinite in QFT). In every tangle process
which we want to calculate we are allowed to remove all such loops by 
multiplying with dim$\,V$. It can be asked, if we can find a Clifford 
biconvolution Hopf gebra which has a unit multiplicative factor. 
This reverses the arguments: {\it physics models the Hopf gebraic axioms},
and not vice versa.

We insist in Hopf gebras, since the antipode seems to be a necessary part
of such a framework. Looking at the antipode axiom, Fig. \ref{antipode}, 
(which might be modified also) the important feature is that the 
output of this tangle is a scalar times identity whatever one throws in. 
This antipode tangle could probably be used as an surrogate for an 
expectation value. Moreover, the antipode splits tangles into parts which
is essential in some proofs.

{\bf Problems:} We have lots of open problems to solve: Find all
basic tangles of a realistic (even free) QFT. Give rules to extract all
loops in such tangles, like Fig. \ref{loop}, to basic loops or non-loops
which can be boxed in basic tangles. Derive from such rules the product,
coproduct and antipode of a Clifford Hopf gebra as a model. Find a 
translation between usual QFT and such a tangle based description, e.g. 
for vacuum expectation values, transition matrix elements, $\ldots$ etc.
\vspace{-0.4truecm}
\section{Acknowledgement}
\vspace{-0.4truecm}
Many fruitful discussions with Prof. Z. Oziewicz during his visit 
in Konstanz have influenced the final version of this work which deviates
considerably from its plan. A travel grant from the DFG is greatfully 
acknowledged.

\vspace{-0.4truecm}
\footnotesize
\begin{thebibliography}{99}
%
\vspace{-0.4truecm}
\bibitem{Caianiello} E.R. Caianiello: \textit{Combinatorics and 
Renormalization in Quantum Field Theory}, W.A. Benjamin, Inc. 1973
%
\bibitem{Con} O. Conradt: \textit{The principle of duality in Clifford
algebra and projective geometry}, in ``Clifford algebras and their
application in mathematical physics'', R. Ab{\l}amowicz,
B. Fauser Eds., Birkh\"auser, Boston 2000, 157--193
%
\bibitem{Cruz} J.J. Cruz Guzman, Z. Oziewicz: \textit{Unital and antipodal
convolution Hopf gebra}, in: ``Miscellanea Algebraica'', W. Korczy\'nski,
A. Obtu{\l}owicz, Eds. Akademia Swietokrzyska, Kielce, submitted 
%
\bibitem{F-vertex} B. Fauser: \textit{Vertex normal-ordering as a
consequence of nonsymmetric bilinear forms in Clifford algebras},
J. Math. Phys. 37(1) 1996, 72--83 
%
\bibitem{F-positronium} B. Fauser, H. Stumpf: \textit{Positronium
as an example of algebraic composite calculations}, in ``The Theory 
of the Electron'' Cuautitlan, Mexico 1995, J. Keller, Z. Oziewicz Eds.,
Adv. Appl. Clifford Alg. 7(suppl.) 1997, 399--418
%
\bibitem{F-thesis} B. Fauser: \textit{Clifford-algebraische 
Formulierung der Quantenfeldtheorie und Regularit\"at}, 
Thesis, Univ. T\"ubingen, 1996
%
\bibitem{F-dirac} B. Fauser: \textit{Dirac theory from a field 
theoretic point of view}, in Aachen 1996, ``Clifford algebras and their 
applications  in mathematical physics'', 
V. Dietrich, K. Habetha, G. Jank Eds., Kluwer, Dordrecht 1998, 89--107
%
\bibitem{F-vacua} B. Fauser: \textit{Clifford geometric parameterization
of inequivalent vacua}, hep-th/9710047, submitted
%
\bibitem{F-transition} B. Fauser: \textit{On an easy transition
from operator dynamics to generating functionals by Clifford algebras},
J. Math. Phys. 39(9) 1998, 4928--4947
%
\bibitem{FauAbl} B. Fauser, R. Ab{\l}amowicz: \textit{On the decomposition
of Clifford algebras of arbitrary bilinear form}, in ``Clifford algebras
and their application in mathematical physics'', R. Ab{\l}amowicz,
B. Fauser Eds., Birkh\"auser, Boston, 2000, 3341--366
%
\bibitem{F-wick} B. Fauser: \textit{On the Hopf algebraic origin of
Wick normal-ordering}, hep-th/0007032, submitted 
%
\bibitem{FauOzi} B. Fauser, Z. Oziewicz: \textit{Clifford Hopf gebra for
two-dimensional space}, in: ``Miscellanea Algebraica'', W. Korczy\'nski,
A. Obtu{\l}owicz, Eds. Akademia Swietokrzyska, Kielce, submitted 
%
\bibitem{GL} R. Grossmann, R.G. Larson: \textit{Hopf algebraic structures
of families of trees}, J. Alg. 26, 1989, 184--210
%
\bibitem{Kreimer} D. Kreimer:\textit{Knots and Feynman diagrams},
Cambridge Univ. Press, Cambridge, 2000
%
\bibitem{Ozi-FGTC} Z. Oziewicz: \textit{From Grassmann to Clifford},
in ``Clifford algebras and their applications in mathematical physics''
J.S.R. Chisholm, A.K. Common Eds., Kluwer, Dordrecht 1986, 245--256
%
\bibitem{Qzi-CAoM} Z. Oziewicz: \textit{Clifford algebras of Multivectors},
Adv. Appl. Clifford Alg. 7(suppl.) 1997, 467--486
%
\bibitem{Ozi-CJP} Z. Oziewicz: \textit{Clifford Hopf gebra and bi-universal
Hopf gebra}, Czech. J. Phys. 47(12) 1997, 1267--1274
%
\bibitem{Ozi-ixtapa} Z. Oziewicz: \textit{Guest Editor´s Note},
Special issue, Int. J. Theor. Phys. Vol. 40(1), 2001, 5 
%
\bibitem{RotSte} G.-C. Rota, J.A. Stein: \textit{Plethystic Hopf algebras},
Proc. Natl. Acad. Sci. USA 91, 1994, 13057--13061
%
\bibitem{StuBor} H. Stumpf, Th. Borne: \textit{Composite particle
dynamics in quantum field theory}, Vieweg Verlag, Braunschweig, 1994
\end{thebibliography}
\end{document}